\documentclass{article}
\usepackage{spconf,amsmath,graphicx,amssymb}
\usepackage{textcomp,siunitx,multicol}
\usepackage[utf8]{inputenc}
\usepackage{verbatim,balance}
\usepackage{amsthm,lipsum,cite}
\usepackage{tikz,siunitx,verbatim}
\usepackage[inline]{enumitem}

\usepackage{pgfplots}
\usepackage[hyphens]{url}
\usepackage[hidelinks]{hyperref}
\pgfplotsset{compat=newest}
\pgfplotsset{plot coordinates/math parser=false}
\newlength\figureheight
\newlength\figurewidth

\usetikzlibrary{shapes,arrows,positioning,decorations.pathreplacing,calc}
\usetikzlibrary{positioning,dsp,calc,shapes.callouts,shapes.arrows}
\usetikzlibrary{mindmap,backgrounds,automata,fit,dsp}

\usepackage{pgfplots}

\definecolor{s1}{RGB}{0,0,0}
\definecolor{s2}{RGB}{130,130,130}
\definecolor{s3}{RGB}{169,169,169}

\definecolor{s4}{RGB}{190,190,190}

\pgfplotscreateplotcyclelist{mylist}{%
{s1,mark=square*,mark options={solid,fill=s1},dashed,mark size=2},
{s2,mark=otimes*,mark options={solid,fill=s2},dotted,mark size=2},
{s3,mark=triangle*,mark options={solid,fill=s1},solid,mark size=2},
{s1,mark=square*,mark options={solid,fill=s2},dashed,mark size=2},
{s2,mark=otimes*,mark options={solid,fill=s1},dotted,mark size=2},
{s3,mark=diamond*,mark options={solid,fill=s3},solid,mark size=2},
{s1,mark=pentagon*,mark options={solid,fill=s2},dashed,mark size=2},
{s2,mark=star,mark options={solid,fill=s1},dotted,mark size=2},
{s3,mark=triangle*,mark options={solid,fill=s3},solid,mark size=2}}

\pgfplotsset{every  axis  legend/.append  style={font=\normalsize}}

\pgfplotsset{tick  label  style={
font=\normalsize}}

\pgfplotsset{every  axis  label/.append  style={font=\normalsize}}

\def\x{{\mathbf x}}

\title{Speech Coding, Speech Interfaces and IoT\\ -- Opportunities and Challenges}
\name{Tom Bäckström\thanks{Supported by the Academy of Finland research project No 312490.}}
\address{Aalto University, Department of Signal Processing and Acoustics, Espoo, Finland}
\hypersetup{breaklinks=true}
\begin{document}

\maketitle
\begin{abstract}
Recent speech and audio coding standards such as 3GPP Enhanced Voice Services match the foreseeable needs and requirements in transmission of speech and audio, when using current transmission infrastructure and applications. Trends in Internet-of-Things technology and development in personal digital assistants (PDAs) however begs us to consider future requirements for speech and audio codecs. The opportunities and challenges are here summarized in three concepts: collaboration, unification and privacy. First, an increasing number of devices will in the future be speech-operated, whereby the ability to focus voice commands to a specific devices becomes essential. We therefore need methods which allows collaboration between devices, such that ambiguities can be resolved. Second, such collaboration can be achieved with a unified and standardized communication protocol between voice-operated devices. To achieve such collaboration protocols, we need to develop distributed speech coding technology for ad-hoc IoT networks. Finally however, collaboration will increase the demand for privacy protection in speech interfaces and it is therefore likely that technologies for supporting privacy and generating trust will be in high demand.
\end{abstract}

\section{Introduction}
A dominant trend in speech processing technology is smart-devices which include
speech interfaces. Personal digital assistants such as Siri~\cite{siri} and
Google~\cite{google} have become
commonplace features of smart-phones, while smart-speakers with speech interfaces
such as Amazon Echo\cite{amazonecho}, Google Home~\cite{googlehome} and Apple
Home~\cite{applehome} have increased their sales greatly~\cite{canalys}.
A central benefit of such speech interfaces is that they provide a unified
hands-free access to services, including news and information retrieval as
well as control of local actuators and devices. Specifically, in comparison
to a tactile interface, requirements to the users physical
proximity to the device is relaxed. This is beneficial especially when accessing
multiple devices in different locations.

In terms of user-interface design, speech interfaces are thus
beneficial in the sense that the user does not need to focus on the location
of devices, but can concentrate on the primary task. For example, if the user wants
to turn of the lights, he can say ``\emph{Computer, lights off}'', even if he does not
know where the computer is located. In contrast, with a tactile interface, the
user first has to locate and reach the physical device, before he can even start
to operate it. Thus with speech interfaces, the user's \emph{attention is shifted from device to task},
which is beneficial since focusing on the device is a distraction with respect to
the task.
Clearly the benefit of such speech interfaces is emphasized when the number of
devices increases.

In a parallel development, Internet of Things (IoT) technologies have recently
received much attention~\cite{miorandi2012internet}. To get the advantage of a
large range of devices, they need to be connected and we need to be able to access
them through a unified interface. For unification, the user interface does not
necessarily have
to be a speech interface, but again, a speech interface relaxes the requirement for
physical proximity in comparison to tactile interfaces. Speech interfaces are thus often the
logical choice for IoT devices.

\begin{figure}
      \centering{\resizebox{\columnwidth}{!}{
    \begin{tikzpicture}[scale=1]
      \node (head) {\includegraphics[width=0.9cm]{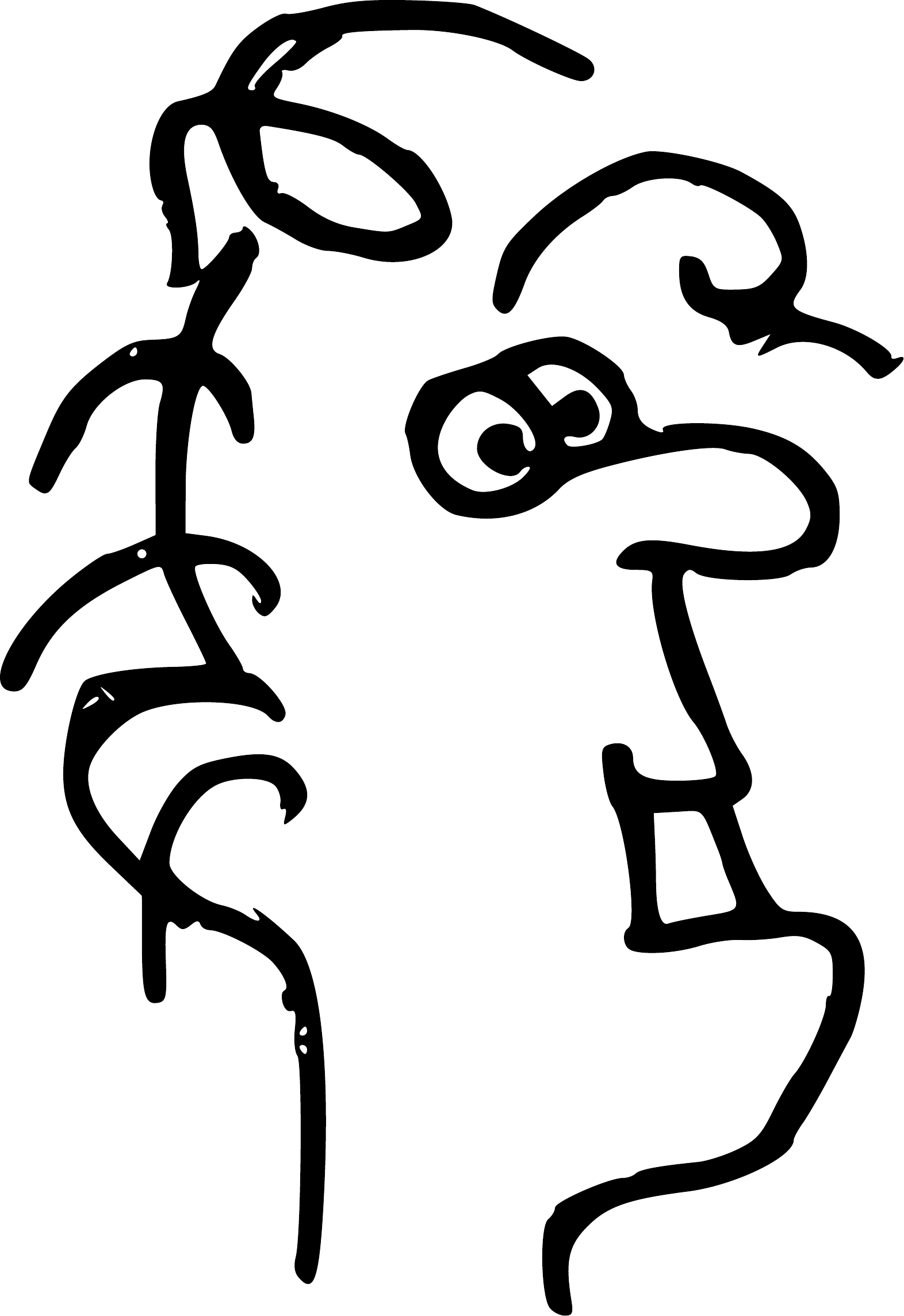}};   
      \draw (0.7,-0.5) arc (-30:30:1);
      \draw (0.65,-0.4) arc (-30:30:.8);
      \draw (0.6,-0.3) arc (-30:30:.6);
      \node[draw,circle,right=of head,xshift=-.4cm,scale=0.5] (mic2) {};
      \node[draw,circle,above=of mic2,scale=0.5] (mic1) {};
      \node[draw,circle,below=of mic2,scale=0.5,yshift=-10mm] (mic3) {};
      \draw[line width=0.2mm] ([yshift=2.5pt]mic1.west) -- ([yshift=-2.5pt]mic1.west);
      \draw[line width=0.2mm] ([yshift=2.5pt]mic2.west) -- ([yshift=-2.5pt]mic2.west);
      \draw[line width=0.2mm] ([yshift=2.5pt]mic3.west) -- ([yshift=-2.5pt]mic3.west);
      \node[draw,right=of mic1,xshift=-.3cm] (enc1) {Encoder};
      \node[draw,right=of mic2,xshift=-.3cm] (enc2) {Encoder};
      \node[draw,right=of mic3,xshift=-.3cm] (enc3) {Encoder};
      \draw[->,very thick] (mic1) -- (enc1);
      \draw[->,very thick] (mic2) -- (enc2);
      \draw[->,very thick] (mic3) -- (enc3);
      \node[draw,right=of enc2.east,yshift=-2mm,xshift=-.3cm,anchor=north,rotate=90,minimum width=4cm,minimum height=.6cm] (tx) {~Transmission channel~};
      \draw[->,very thick] (enc1) -- ($(enc1.east)+(.7cm,0)$);
      \draw[->,very thick] (enc2) -- ($(enc2.east)+(.7cm,0)$);
      \draw[->,very thick] (enc3) -- ($(enc3.east)+(.7cm,0)$);
      \node[draw] at ($(enc1.east)+(2.75cm,0)$) (dec1) {Decoder};
      \node[draw] at ($(enc2.east)+(2.75cm,0)$) (dec2) {Decoder};
      \node[draw] at ($(enc3.east)+(2.75cm,0)$) (dec3) {Decoder};
      \draw[->,very thick] ($(dec1.west)+(-.7cm,0)$) -- (dec1);
      \draw[->,very thick] ($(dec2.west)+(-.7cm,0)$) -- (dec2);
      \draw[->,very thick] ($(dec3.west)+(-.7cm,0)$) -- (dec3);
      \node[draw,right=of dec2] (merge) {\parbox{1.4cm}{Channel merge}};
      \draw[->,very thick] (dec1.east) -- (merge);
      \draw[->,very thick] (dec2.east) -- (merge);
      \draw[->,very thick] (dec3.east) -- (merge);
      \node[]  at ($(merge.east)+(1cm,0)$) (loudspeaker2) {\includegraphics[width=0.6cm]{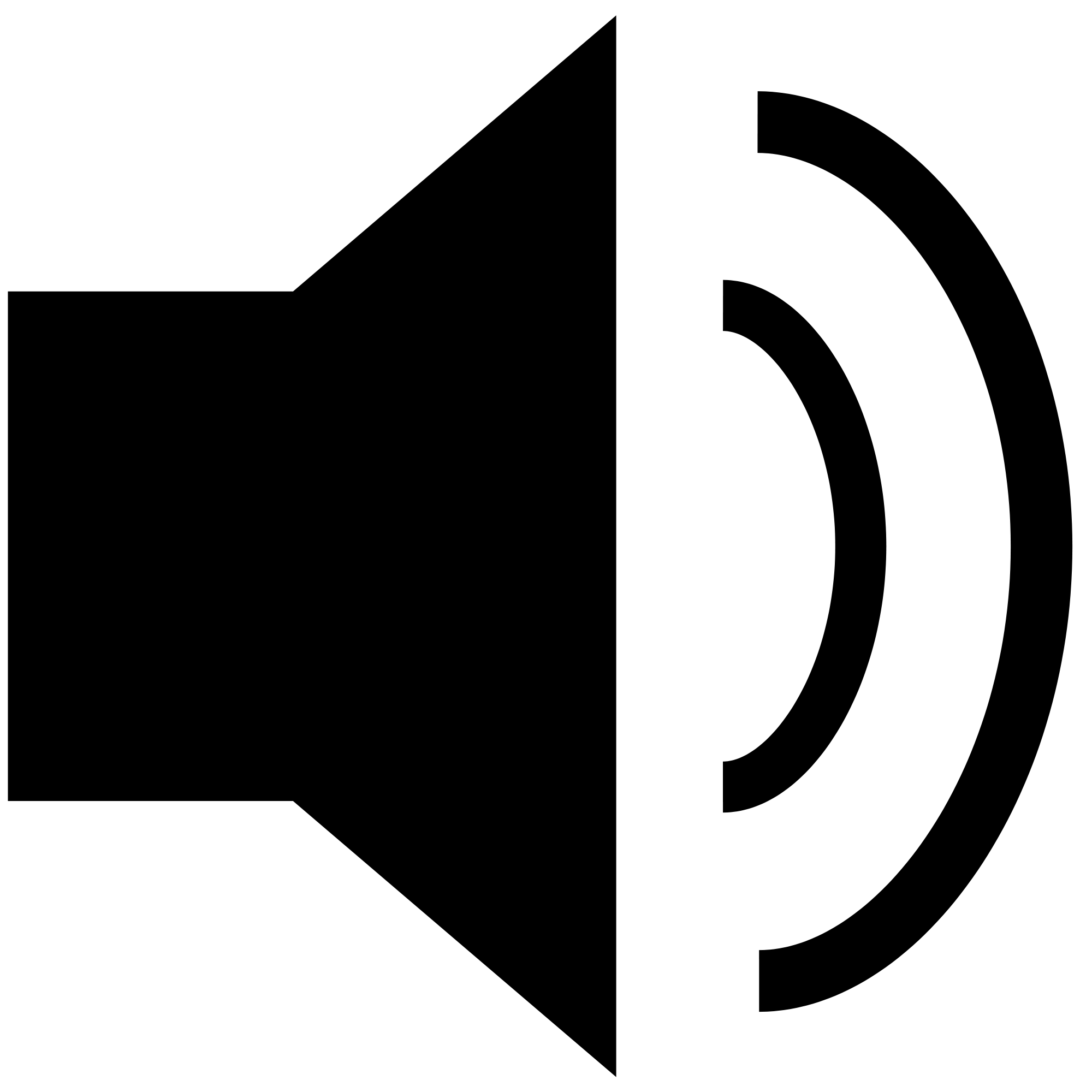}};
      \draw[->,very thick] (merge) -- (loudspeaker2);
      \node[right=of loudspeaker2,xshift=-1cm] (listener2) {\includegraphics[width=1.2cm]{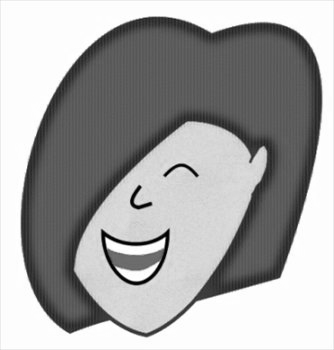}};
      \draw[dotted,thick] ($(enc2.south)+(0,-2mm)$) -- ($(enc3.north)+(0,2mm)$);
      \draw[dotted,thick] ($(dec2.south)+(0,-2mm)$) -- ($(dec3.north)+(0,2mm)$);
      \node[draw,dashed,minimum height=4cm,minimum width=4.7cm] at ($(dec2.east)+(5mm,-2mm)$) {~};
      \node[above=of dec1,xshift=1cm,yshift=-.5cm] {Merge node};
      \node[above=of enc1,xshift=-.5cm,yshift=-.5cm] {Microphone nodes};
        \end{tikzpicture}}

      }

  \caption{Illustration of a distributed acoustic sensor network.}\label{fg:aswn}
  
\end{figure}
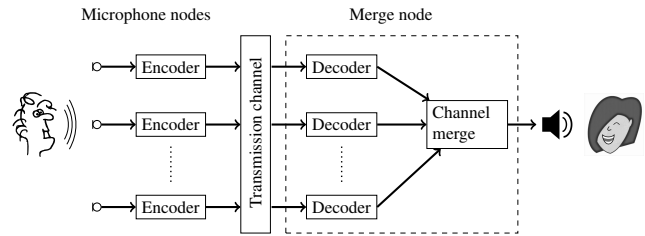

Further significant benefits of IoT devices can be achieved with devices 
equipped with microphones. A collection of devices can then act like a (wireless) acoustic
sensor network (WASN), to provide efficient pick-up of acoustic signals even in adverse
conditions~\cite{bertrand2011applications} (see Fig.~\ref{fg:aswn}).
Specifically, it is possible to apply multi-microphone enhancement
methods such as noise attenuation, beamforming and source separation,
to extract individual spatially separated sound
sources~\cite{benesty2008springer}.

Communication between such devices requires speech and audio coding techniques,
which is a field which has recently seen several advances. Namely, the 3GPP has
standardized the Enhanced Voice Services (EVS) for telecommunications, while MPEG has
standardized Unified Speech and Audio coding for speech and audio broadcasting, as well as
3D-Audio for multichannel broadcasting, and at the time of writing, Bluetooth is about to
publish their new codec for low-delay speech and audio
communication~\cite{3GPP2014,MPEG-D-USAC:2012,herre2015mpeg,backstrom2017celp}.
These codecs are state-of-the-art technology and provide the best known compromise between
quality and available resources. However, none of these codecs are suitable for the IoT
scenario. The available codecs are for communication between single devices, whereas for
IoT, we need simultaneous communication between multiple devices.

In communication between multiple independent devices in ad-hoc networks, conventional
single-device codecs give a suboptimal performance, because each device can then in theory
be sending the same signal, which is redundant and thus inefficient. Methods of distributed source
coding can then be used to avoid such redundancy~\cite{xiong2004distributed,dong2006distributed,
  barriac2004distributed,jia2006distributed,majumdar2003distributed}. It
is then in principle possible to achieve optimal compromise between rate and distortion, while
simultaneously optimizing resource consumption between devices. Available methods for distributed
source coding have however not been able to provide competitive performance in a single channel
setting. 

In contrast, we have recently presented methods which extend single device codecs with features
of distributed coding, with the objective of improving quality when multiple devices are available,
that is, when we can use distributed coding~\cite{backstrom2016iot,backstrom2018dithered,backstrom2018fastrand}.
The initial approach here is to decorrelate quantization error between devices, such that they
can be removed by post-filtering methods~\cite{das2018postfiltering_cplx,das2018postfiltering_log}.
While our current methods do not yet reach the full benefit of distributed source coding, they do retain
competitive quality in single-device scenarios while enabling additional devices to
improve quality.\smallskip

Given this \emph{scientific} background, the purpose of the current work is to highlight some
problems in state-of-the-art \emph{implementations} in commercial products, and to propose a
path to resolving these issues to the benefit of all actors. In short, this paper argues for a
standard communication protocol between IoT devices, a unified speech and audio codec for both telecommunication
and digital services, which is designed to ensure the privacy of the users. Indeed, the presented arguments
point to a system where minimization of the amount data transmitted provides best resource efficiency,
minimizes hardware requirements and provides best privacy.

The challenges with state-of-the-art implementations are described through two use cases, the cafeteria and the bedroom, described
in the following section, which have been chosen to represent
typical scenarios for a
large percentage of the users. As a solution to such problems, in Section~\ref{sec:local}, we propose a
systems design where devices are allowed to collaborate when in the same acoustic space.
In the discussion section we further comment also on the ethical and  legal landscape of speech processing
in multi-device scenarios.

\section{Use cases}
\subsection{Cafeteria}

Suppose two friends, Alice and Bob, meet at their favorite cafeteria. Alice happens to prefer products
from the Apple brand, while Bob uses an Android phone. The cafeteria also provides access to
Amazon Echo devices. Alice and Bob then have access to \emph{three} different speech interfaces. Which
interface should they use?

I am in no position to prescribe preference among products, but the point is that Alice and Bob are
\emph{forced to focus on devices} to choose their speech interface. In contrast, a central promise
of speech interfaces was to shift attention away from devices to the task at hand. Current products
give no solution to that issue. If Alice and Bob want to
call their mutual friend Steven, choosing a device for that task is  a distraction, which reduces
their user experience.

This unsatisfactory experience occurs because current products which provide speech interfaces are
\emph{not interoperable} across manufacturers and service providers. Though the argument for speech interfaces
was centered
around the promise for a shift of attention from device to task, current products are unable to fulfill this promise
because users have to choose the device and service jointly.
Specifically, the available products are stellar products when observed in isolation, but when devices
of multiple manufacturers are used in parallel, performance is unsatisfactory.

Conversely, to achieve the shift of attention from device to task, devices should collaborate, which means that
they should be interoperable. Collaboration across different manufacturers is probably best achieved
with a common application programming interface (API). To maximize the number of manufacturers which support
such an API, it seems obvious that the API should be open or at least free-to-use. Several different
standardization organizations already have the necessary procedures for adopting such technology, whereby
standardization is a viable alternative.

\subsection{Bedroom}
When I am in my bedroom and say ``\emph{Computer, lights off},'' the lights are turned off. This is an
extremely private event; it provides insight into my habits in the bedroom. Nobody, with the exception
of my preferred romantic partner, has access to my bedroom, not even my best friends. It is therefore
inconceivable to me how someone would provide access to his or her bedroom to some company providing
a cloud service.

I would personally contend that the privacy provided by current products is not a happy state of affairs,
but objectively, increasing collaboration across devices between different manufacturers, will increase
privacy concerns exponentially. Speech interfaces are often served through cloud servers and if all
speech input would
be shared among the cloud servers of all nearby devices, the weakest link would define the level of privacy
of all users. It would be very hard for the user to determine what that the actual level of privacy is and
consequently, it would be extremely hard for the user to trust the system.

Privacy becomes a concern when information is transmitted or stored. In the most basic form of privacy
protection, users need to be able to protect their own information to be transmitted or stored without
their consent. Some of the main issues are that users often do not know what information is transmitted,
to whom, or how information can be used. Moreover, companies which provide services and manufacture devices
generally have more expertise than the average user, whereby they companies both know better what information
is valuable but also may know how to extract information without alerting the average users' suspicion.

Among privacy and security advocates it is widely agreed that the main solution cannot be education of
users~\cite{schneier2016stop}. Companies and potential criminals always have superior expertise compared
to average users and it is unreasonable to expect that we could educate users to such an extent that the
difference in expertise
would be overcome. Instead, technology should be designed to protect the users. Collaboration among speech
interfaces should be designed such that it minimizes information transfer to the minimum amount 
required to fulfill the task at hand. In particular, to follow an ethically aligned design
paradigm~\cite{ieee2017ethically}, services and information transfer should be activated only through
informed and meaningful consent.

\section{Local Collaboration}\label{sec:local}
\begin{figure}
    \centering\resizebox{.75\columnwidth}{!}{
  \begin{tikzpicture}[scale=1.2]
    \node[minimum width=1cm] at (0,0) {\includegraphics[width=1.4cm]{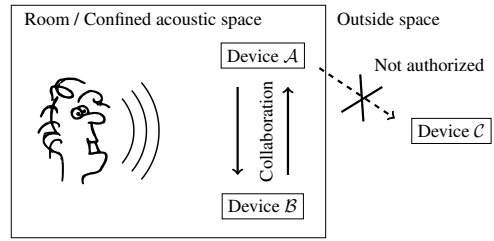}};
    \draw[thick,domain=-40:40]  plot({0.15+0.8*cos(\x)}, {0.8*sin(\x)});
    \draw[thick,domain=-40:40]  plot({0.15+1.0*cos(\x)}, {1.0*sin(\x)});
    \draw[thick,domain=-40:40]  plot({0.15+1.2*cos(\x)}, {1.2*sin(\x)});
    \node[minimum width=1.5cm,draw] at (3,1.2) (A) {Device $\mathcal{A}$};
    \node[minimum width=1.5cm,draw] at (3,-1.2) (B) {Device $\mathcal{B}$};
    \draw[->,very thick] (2.6,0.7) -- (2.6,-0.7);
    \draw[<-,very thick] (3.4,0.7) -- node[above,rotate=90,yshift=0.15cm] {Collaboration} (3.4,-0.7);
    \draw (-1,-1.7) -- (-1,2) -- (4,2) -- (4,-1.7) -- (-1,-1.7);
    \node at (1.1,1.75) {Room / Confined acoustic space};
    \node at (5.0,1.75) {Outside space};
    \node[minimum width=1.5cm,draw] at (6,0) (C) {Device $\mathcal{C}$};
    \draw[->,very thick,dashed] (3.9,1.0) --  (5.1,0.2);
    \draw[very thick] (4.25,0.3) -- (4.9,0.7);
    \draw[very thick] (4.6,0.15) -- (4.55,0.9);
    \node at (5.65,1.05) {Not authorized};
  \end{tikzpicture}}

    \caption{Illustration of local collaboration between devices in the same acoustic space,
    where connections to the outside are forbidden without specific authorization.}
    \label{fg:local}
  
\end{figure}

A solution which fulfills the above requirements is a systems structure where only local collaboration
is permitted by default (see Fig.~\ref{fg:local}). All devices which are within earshot, that is, which
can receive my speech with their microphones, already have access to the speech signal. Collaboration
between such local devices, $\mathcal A$ and $\mathcal B$, does thus not expose the user to 
new threats to privacy. Only remote devices $\mathcal C$, remote in space or time, expose the user's
privacy, whereby transmission to such devices require separate and specific authorization.
By remote in space or time, respectively, we refer to telecommunications and storage applications.

The benefit of local collaboration is that it allows resource-optimization among local devices.
Each device could in theory run its own noise attenuation, source separation, voice activity detection,
keyword spotting and speech recognition algorithms and more, but those algorithms would then be redundant,
since multiple devices run the same tasks. For improved efficiency and accuracy, we can instead use, for example,
a distributed voice activity detector among local devices~\cite{berisha2006real}, distributed beamforming~\cite{barriac2004distributed} or distributed speech recognition~\cite{etsi2003distrec}.

It is here evident that if a local device, such as $\mathcal A$ and $\mathcal B$, has been compromised
to allow access to an illegitimate user, then that device can leak microphone signals to the outside. The current
proposal provides no tools for avoiding such leaks. Device-level security however falls outside the current
scope and in contrast, this proposal is only concerned with information flows between legitimate users and
services.

Specifically, for example at home, my family members have access to the same network and we trust each other.
However, if I go into the study and close the room behind me, that is a clear sign that I want to have privacy.
Whereas all the devices of family members are generally allowed to collaborate, when I close the door behind me,
I limit access to only those devices which are with me in the same room.

To achieve such functionality, devices have to be able to sense whether they are in the same room or
acoustic space, and such sensing must not reveal any private information. Since the objective is to
determine access rights to an acoustic signal, it is easiest to use the acoustic signal also as the basis
of access right management. Specifically, each device can extract data from the microphone signal,
apply error correction to obtain a fingerprint which is associated to that unique acoustic environment
and use that fingerprint to cipher communication among local devices~\cite{schurmann2013secure}.
Since the fingerprint is used as a key for encryption, it is never transmitted between devices and no sensitive
data is exchanged unencrypted. Moreover, since only such devices which reside in the same acoustic
space can know the key, only local devices can decrypt the messages and participate in the collaboration.

\begin{figure}
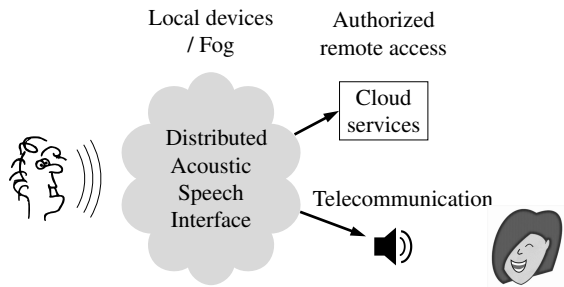

   {\centering\resizebox{.9\columnwidth}{!}{
  \begin{tikzpicture}[scale=1.5,node distance=3mm]
    \node (head) {\includegraphics[width=0.9cm]{Cartoon-Head}};   
    \draw (0.54,-0.4) arc (-30:30:.8);
    \draw (0.47,-0.35) arc (-30:30:.7);
    \draw (0.4,-0.3) arc (-30:30:.6);
      \node[overlay,cloud,aspect=.8,fill=gray!30] at ($(head.east)+(1.4cm,0)$) (cloud) {\parbox{1.4cm}{\centering Distributed Acoustic Speech\\ Interface}};
      \node at ($(head.east)+(1.4cm,1.5cm)$) {\parbox{2cm}{\centering Local devices / Fog}};
      \node at ($(cloud.east)+(1cm,1.5cm)$) {\parbox{2cm}{\centering Authorized\\ remote access}};
      \node at ($(cloud.east)+(1.2cm,-.2cm)$) {Telecommunication};
      \node[draw,align=center] at ($(cloud.east)+(1cm,.7cm)$) (siri2) {Cloud\\ services};
      \draw[dspconn,->,very thick] (cloud) -- (siri2.west);

      \node[xshift=6mm]  at ($(cloud.east)+(.7cm,-.7cm)$) (loudspeaker2) {\includegraphics[width=0.6cm]{Loudspeaker}};
      \draw[dspconn,->,very thick] (cloud) -- (loudspeaker2);
      \node[right=of loudspeaker2,xshift=6mm] (listener2) {\includegraphics[width=1.2cm]{happy-lady-bw}};
     \end{tikzpicture}  }

   }
   \caption{Illustration of the proposed systems design for a distributed acoustic speech interface
   acting as an abstraction layer to unify access to local devices.}\label{fg:fog}
  
\end{figure}

\section{Systems design}

Conceptually, the proposed functionality can be interpreted as a new abstraction layer, which lumps together
all local devices, to only give out the jointly estimated speech signal (see~Fig.~\ref{fg:fog}). Such an
abstraction layer comprising local devices can be named the \emph{fog} as a parallel to the abstraction of
remote servers as a \emph{cloud}. Remote services do not need to know the particulars of local devices --
the cloud needs access to only the output sound -- whereby it is well-warranted to hide such information
from the cloud. Any local device can then act as a merge-node where the desired speech signal is estimated
from distributed sensor nodes (see Fig.~\ref{fg:aswn}).

To transmit speech to remote recipients then requires that local devices reach a consensus on
authorization~\cite{coulouris2005distributed}. Such authorization mechanisms must prevent spoofing attacks
from any single device which implies that the authorization mechanism should be distributed and immutable.
In other words, authorization likely requires distributed ledger technologies, which are colloquially known
as block-chain technologies~\cite{deshpande2017distributed}.

\section{Discussion and Conclusions}
The presented use-cases lead to the conclusion that, for voices services in IoT devices, it would be
beneficial to develop a new speech and audio coding standard applicable to multi-device scenarios. In comparison
to prior standards, the novelty of such a standard would be that it incorporates methods of both distributed
source coding and built-in provisions for protecting the privacy of its users. Central benefits of such
a standard include that
\begin{enumerate*}
\item users can access services and devices without the necessity of physical proximity to the
  device, thus releasing attention from the device to the task at hand,
\item microphones in several devices can be used to jointly estimate the speech signal for better signal
  quality,
\item available hardware and software resources can be optimized to minimize costs in infrastructure and
  energy consumption and
\item a unified and transparent approach to privacy helps to build trust to technology among users as well
  as allows third-party audits of privacy.
\end{enumerate*}

Recent work on privacy has highlighted the need for the users to be able to manage their own
information~\cite{kuikkaniemi2014mydata} and the general data protection regulation recently legislated
by the European Union~\cite{eu2016gdpr} enforces such a model in practice. In comparison, speech information
is by nature a dialogue between two or more parties and it seems intuitively clear that ownership of and
access rights to such a
dialogue is always shared among the participants. The proposed system takes a first step in that direction
by enabling a distributed access management system, where access to data is granted through a consensus among
participants.

This approach to privacy is in stark difference to current products, where access management is in some
cases governed by a single primary user who owns the device. This exposes secondary users to a breach
in trust. For example, if I have a smart speaker
at home and I could  remotely access the voice command history, then while away from home, I could listen
to the commands other family members make at home. Though I do not have the expertise to determine whether
this is legal, certainly this situation seems ethically unreasonable.

In conclusion, smart speakers and other smart devices have recently become commonplace products and they
provide useful hands-free access to services through voice operated user-interfaces. The current conclusion
is that the usability, signal quality and resource efficiency would improve if all IoT devices could collaborate
and jointly provide a speech user-interface. The proposed approach is a unified multi-device speech and audio
codec for telecommunication and digital services. An essential part of such a codec would be provisions for
enforcing and maximizing users' privacy.

\bibliographystyle{IEEEbib}
 \newcommand{\noop}[1]{}

\balance

\end{document}